# Threshold concentration for H blistering in defect free W

W. Xiao,[1] G. -N. Luo,[2] and W. T. Geng[1a]

*School of Materials Science & Engineering, University of Science & Technology Beijing, Beijing 100083, China,*

*Institute of Plasma Physics, Chinese academy of Sciences, Hefei 230031, China*

January 13, 2011

Lattice distortion induced by high concentration of H is believed to be precursor of H blistering in single crystalline W (SCW) during H isotope irradiation. However, the critical H concentration needed to trigger bond-breaking of metal atoms presents a challenge to measure. Using density functional theory, we have calculated the formation energy of a vacancy and a self-interstitial atom (SIA) in supersaturated defect-free SCW with various H concentrations. When the ratio of H:W exceeds 1:2, the formation of both vacancies and self-interstitials becomes exothermic, meaning that spontaneous formation of micro-voids which can accommodate molecular $H_2$ will occur. Molecular $H_2$ is not allowed to form, and it is not needed either at the very initial stage of H blistering in SCW. With supersaturated H, the free volume at the vacancy or SIA is greatly smeared out with severe lattice distortion and more H can be trapped than in the dilute H case.

[a] To whom correspondence should be addressed. E-mail: geng@ustb.edu.cn

The hydrogen implanted into plasma-facing materials can form bubbles, posing a great threat to the stability of a fusion reactor. Serving as promising candidate armor material for some parts of the divertor and first wall in future reactors, tungsten has been under intensive study for its response to H isotope irradiation in recent years. [1,2,3,4,5] Elucidation of the relationship between material properties such as diffusivity, solubility, permeability, and recombination coefficient of H to the irradiation conditions such as temperature, energy, flux, and fluence is a must in material development. The H blistering in polycrystalline W is generally believed to start with retention and accumulation of H at intrinsic defects such as vacancies, grain boundaries, and dislocations, or extrinsic defects like impurities. [6,7,8] In the case of single crystalline W (SCW), a much higher fluence of implantation or lower temperature is needed to trigger H blistering, [9,10,11,12] prompting some authors to propose that it is the ultra-high concentration of H that causes high stresses, a release of which assists starting phase separation of H and W possibly through formation of voids and vacancy clusters. [4,13] In either case, a high local concentration of H is needed to initiate blistering.

At present, there are mainly two models proposed to understand H blistering in W, namely, the *gas-driven* model and the *lateral-stress* model. [14,15,16,17] While the *gas-driven* and *lateral-stress* mechanisms for H blistering cannot be easily distinguished and verified in experiment, they clearly differs in the point whether $H_2$ molecules form before cleavage of W-W bonding is activated at the very initial stage. Using density functional theory calculations, Ohsawa *et al*. [18] have shown that a vacancy can trap up to 12 H atoms and no $H_2$ molecule can be formed. In almost simultaneous works, Heinola *et al.* [19]

and Johnson and Carter,[20] have taken into account the temperature factor, and demonstrated that a vacancy only adsorbs five H atoms at room temperature and no molecules form as expected. For another point defect, substitutional He, Jiang *et al.*'s work[21] have demonstrated that it can also attract as many as 12 H atoms and no molecules form. In our recent work,[22] we have shown by first-principles density functional theory calculations that no $H_2$ molecules can form at the intrinsic trapping center in W, including grain boundary, grain boundary combined with vacancy and dislocation loop. As a result, if there are no pre-existed micro-voids, gaseous $H_2$ is unlikely to form prior to blistering during non-damaging irradiation. Regarding the *lateral-stress* model, one point has been missing in previous discussions is that the mobile concentration of implanted H is spatially continuous and the concentration gradient (no more than 1% per lattice constant) is not very large viewed in scale of the crystal lattice. This means that during non-damaging irradiation, the lattice mismatch between neighboring atomic layers in the implantation zone (micrometers away from the surface)[6,7] shall be considerably small. It is therefore quite doubtful that lattice mismatch will generate rippling near the surface.

Here we propose another mechanism for the initiation of blistering. When the concentration of H exceeds a critical value, formation of both vacancies and SIAs becomes exothermic. It follows immediately the spontaneous formation of vacancy clusters or micro-voids, and then $H_2$ molecules will appear at the same time. The *gas-driven* mechanism will then take over. To determine such a critical concentration, we have carried out a DFT study on the formation energy of vacancy and SIA for both pure and a couple of H-supersaturated W systems.

The supercell we employed to calculate the formation energy of a vacancy and SIA contains 128 lattice sites (4×4×4) of bcc W, the same as the one used in literature. Such a size of supercell was shown to yield reliable formation energy for vacancy, SIA, or impurities in dilute cases, with minimized effect of periodic boundary condition.[19,23] Our first-principles total energy DFT calculations were carried out using Vienna *Ab-initio* Simulation Package (VASP).[24] The electron-ion interaction was described using projector augmented wave (PAW)[25,26] potentials, the exchange-correlation between electrons using the generalized gradient approximation (GGA) in the Perdew-Burke-Ernzerhof (PBE)[27] form. We expanded the one-electron wave functions in a plane wave basis with an energy cutoff of 250 eV. The Brillouin-zone integration was performed within Monkhorst-Pack scheme using a (3×3×3) mesh. To obtain the total binding energy for each defective system, the atomic positions were fully optimized with the volume of the supercell being fixed. The lattice constant of the bulk bcc W was calculated to be 3.171 Å, in good agreement with both the experimental value (3.16 Å)[28] and previous DFT results.

In dilute case, the stable position for H in W is the tetrahedral interstitial site.[29] Its solution energy in W, in reference to an isolated $H_2$ molecule, is about 0.96 eV. Therefore, the solubility of H in W is very low, only 1 appm at 1400 K.[3] In the present work, we have investigated four H concentrations, namely, the H:W ratio $c$ as 0.00 (0/128), 0.01 (1/128), 0.50 (64/128), and 1.00 (128/128). For the dilute case ($c$=0.01), we did not optimize the volume of the supercell, but rather kept it as that for the pure W, in view of the fact that H has a much smaller atomic size than W. At very high concentration, however, H will expand remarkably the lattice of W. Early first-principles calculations by Henriksson *et al.*[28] showed that the H-H interaction is strongly repulsive at small distances and a H-H pair in W has very shallow minimum (-0.1 eV) in their binding energy as a function of separation (at 2.2 Å), suggesting that H distribution at high concentration is most likely homogeneous. For this reason, we have set the initial positioning of implanted H atoms with cubic symmetry rather than randomly to save computational effort. For $c$ =0.50, there is one H in the smallest cubic cell of W. Our calculations determined the most stable site for H is the tetrahedral interstitial site, similar to the dilute case. The optimized lattice constant of H-implanted W is now 3.268 Å. To find the right position of the second H for $c$ =1.00, we have examined all the other tetrahedral and octahedral interstitial sites and find that the two H atoms prefer to stay in two tetrahedral interstitial sites separated by 2.36 Å (Figure 1). The optimized lattice constant increased to 3.344 Å. Then this primitive cubic cell was extended by (4×4×4) to serve as the supercell used to calculate the formation energy of vacancy and SIA. To get these energies, we used a chemical potential of W in perfect crystal, in view of the fact high concentration of H occurs only in a very limited region within the implantation zone.

**FIG.1 Computation model of H-implanted W. (a) H:W = 1:2; (b) H:W=1:1**

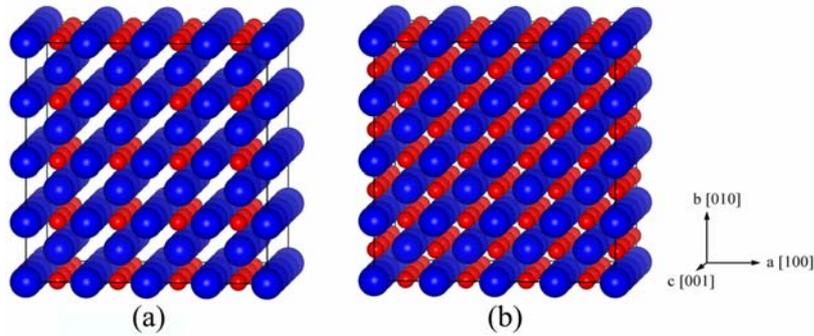

We display in Figure 2 the calculated formation energies for both vacancy and SIA in W, implanted with different amount of H. In the absence of H, the formation energy for a vacancy in W is calculated to be 3.33 eV, in good agreement with previous DFT results.[30,31] As for SIA, the calculated formation energy is 10.28 eV. We have compared the stability of two configurations, in both of which the SIA forms a dumbell with its neighboring W atom. It is found that the alignment with the dumbell in parallel to the <111> direction is 0.28 eV more stable than the one along <110>, confirming Nguyen-Manh's first principles results.[23]

In the dilute case (H:W=1:128), the trapping energy of a vacancy to H, which is the energy release when an interstitial H atom moves from far away to the vicinity of a vacancy, is about -1.3 eV.[18,19,20] Our calculations show that the trapping of H from an SIA is -0.39 eV. This means that with a H atom nearby, the formation energy of a vacancy reduced from 3.33 eV to 2.02 eV. As for SIA, it reduced from 10.28 eV to 9.89 eV. As the H concentration in W rises up to 0.50, the formation energy for a vacancy is decreased to only 0.25 eV, and that for SIA to 3.90 eV. If the H concentration is increased further to 1.00, formation of both vacancy and SIA become strongly exothermic, a condition sufficient for spontaneous formation of vacancies, vacancy clusters, and micro-voids. We note that a precise description of H in metals very often involves the evaluation of zero-point energy (ZPE). The calculated ZPE of H in a bulk interstitial site in W is about 0.26 eV and the ZPE correction to the trapping energy of H atoms to a vacancy is about 0.1-0.2 eV [Ref. 18-21]. Since, as seen above, we are dealing with changes in defect formation energy of several eV in the present work, the ZPE can be safely neglected.

**FIG.2 The formation energy of vacancy (*V*) and self-interstitial (*SIA*) in W with different concentrations of implanted H.**

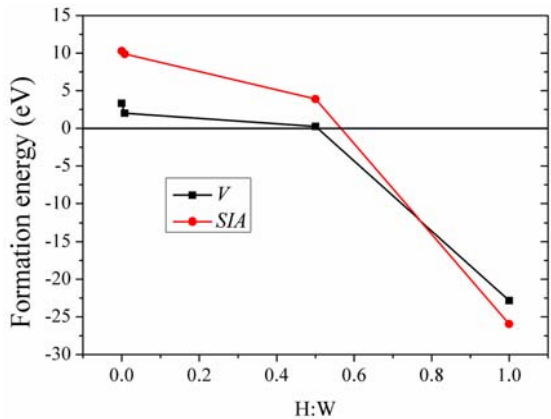

We note that the exact critical concentration of H at which point spontaneous formation of vacancies and SIAs in defect-free single crystalline W start is hard to determine using the model adopted in this work. However, that point is surely above but not far above $c$=0.50.

There are two striking features of the data shown in Fig.2. First, the formation energy of both vacancy and SIA experience a steep drop from $c$=0.50 to $c$=1.00. The calculated solution energies for H in these two cases, with respect to an isolated $H_2$ molecule, are 0.95 and 0.83 eV respectively; whereas in the dilute case, it is 0.96 eV. Thus, H atoms in W are highly unstable and phase separation (bubble formation or diffusion out of the materials) has strong tendency to occur. With higher H concentration, the effect of a vacancy to release the stress through H segregation gets stronger. However, the abrupt drop from $c=0.50$ to $c=1.00$ probably means a dramatic change of the local structure around the vacancy. In Fig.3, we draw the calculated valence charge density on the *a-c* plane (see Fig.1). It is very clear that the removal of one W atom causes a much more significant disturbance of the surrounding H and W in panel (b) than in (a). Associated with the collapse of the local structure, the vacancy is decomposed into several parts, which are evidently illustrated by charge depletion. Since decomposition of a mono vacancy into several sub-vacancies increased the area of low charge-density surfaces, it is much more effective in attracting (trapping) the surrounding H atoms.

**FIG. 3 Charge density on *a-c* plane near a vacancy in W with super-saturated H for concentrations of 0.50 (a) and 1.00 (b). Contours start from $9^{-2}$ e/a.u.$^3$ and increase successively by a factor of $9^{1/9}$.**

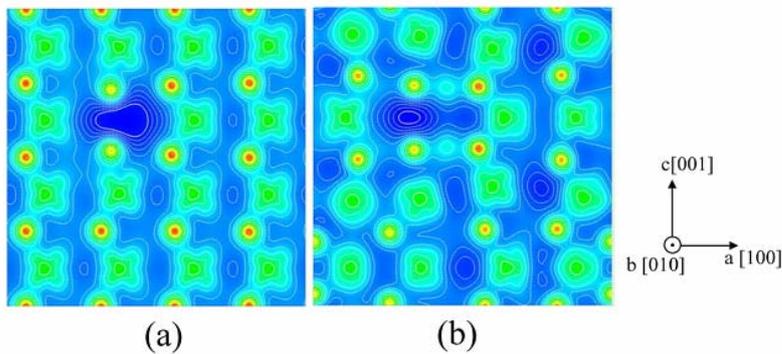

On the other hand, a direct viewing of the SIA's effect is not as illustrative, because of the low symmetry of the local structure. Instead, we resort to the distribution of the bond lengths (nearest-neighbor distances) in analyzing the lattice distortion. In Fig. 4, we plotted the number of H-H (a), H-W (b), and W-W (c) bonds or nearest-neighbor distances in the computation cell within each length range at a particular step for both $c$=0.50 to $c$=1.00, both vacancy and SIA cases. At $c$=0.50, there are sharp peaks of distribution, indicating both H-H (panel a) and H-W (panel b) nearest-neighbor distances stay close to their value in non-defective environment, i.e., 3.27 and 1.88 Å. For $c$=1.00, however, remarkable dispersion occurs near a vacancy or SIA, an indication of redistribution of H in W. We emphasize that no $H_2$ molecules (bond length=0.75 Å) are found for both concentrations. In panel (c), we can see that in the absence of H, there is only one peak of the W-W bond length distribution in both vacancy and SIA cases. At $c$=0.50, another peak emerges. About half W-W bonds have lengths around 2.8 Å, and half around 3.0 Å. This is an artifact of the model we used. As can be seen in Fig. 1, there is one H layer [(100) plane] in between every two W layers. Therefore, W-W bonds across the (100) plane will be slightly different from those in the plane upon lattice expansion. Although there appears a small number of W-W bonds shorter than 2.8 Å or longer than 3.0 Å, the regularity of lattice sites remains. However, at an even higher H concentration $c$=1.00, both vacancy and SIA enhance greatly the dispersion of W-W bond-lengths distribution. As a matter of fact, first and second neighbor for W atoms cannot be distinguished. This means that the W lattice has been drastically distorted.

**FIG. 4 The distribution of H-H (a), H-W (b) and W-W (c) bond lengths (nearest-neighbor distance) in the computation cell for various systems. The length step is 0.1 Å for H-H and W-W, and 0.02 Å for H-W bonds.**

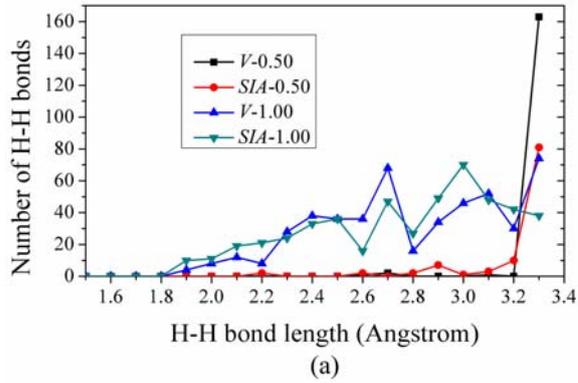

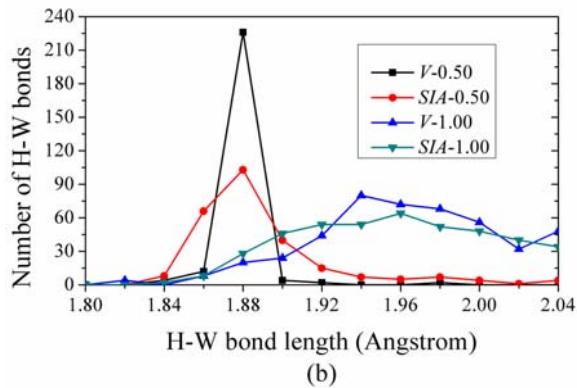

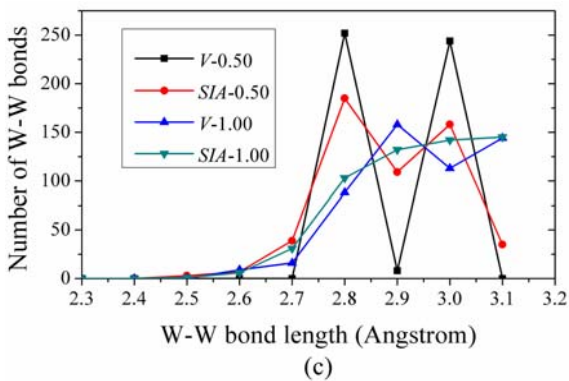

The second striking feature shown by Fig.2 is that the formation of an SIA is even easier than a vacancy at $c=1.00$. This is in sharp contrast to the general cases of compounds.

From Fig. 4c we can find for $c$=1.00, there are fewer short W-W bonds ($\leq$ 2.6 Å) near the SIA (six) than in the vicinity of a vacancy (nine), contrary to the case of $c$=0.50. This is a strong indication that at ultra-high H concentration, the steric effect of an SIA vanishes completely. The resultant extra interstices (with respect to the perfect W-H system) exert similar and even a bit stronger attraction to the surrounding H atoms. Obviously, the spontaneous formation of vacancy and SIA will trigger H blistering. Although $H_2$ molecules will eventually form when the vacancy clusters are big enough, they do not appear at the very beginning of the local lattice collapse in W.

We expect the mechanism of H blistering in single crystalline W illustrated in the present work to work also for other metals. Therefore, our density functional theory result merits experimental scrutiny even though the determination of threshold concentration for H blistering in single crystals by *in situ* measurements presents a big challenge.

**ACKNOWLEDGMENTS**

We are grateful to the support of the NSFC (Grant No. 50971029) and MOST (Grant No. 2009GB109004) of China, and NSFC-ANR (Grant No. 51061130558).